\documentclass[]{svmult}
\pdfoutput=1
\usepackage[bottom=2.5cm,top=2.5cm,bindingoffset=0.8cm,left=1.5cm,right=1.5cm,a5paper]{geometry}
\usepackage{blindtext}
\usepackage{mathptmx}       
\usepackage{helvet}         
\usepackage{courier}        
\usepackage{type1cm}        
\usepackage{makeidx}         
\usepackage{graphicx}        
\usepackage{multicol}        
\usepackage{hyperref}        
\usepackage[bottom]{footmisc}
\usepackage{wasysym}

\usepackage{color}

\makeindex             

\begin{document}

\title*{The Nice Cube (Nice$^3$) nanosatellite project}
\author{F. Millour, S. Ottogalli, M. Maamri, A. Stibbe, F. Ferrero, L. Rolland, S.~Rebeyrolle, A. Marcotto,
K.~Agabi, M. Beaulieu, M. Benabdesselam, J.-B.~Caillau, F. Cauneau, L.~Deneire, F. Mady, D. Mary, A. Memin, G.~Metris, J.-B.~Pomet, O. Preis, R.~Staraj, 
E.~Ait~Lachgar, D. Baltazar, B.~Gao, M. Deroo, B. Gieudes, M. Jiang, T. Livio de Miranda Pinto Filho, M.~Languery, O. Petiot, A.~Thevenon}

\authorrunning{F. Millour, S. Ottogalli et al.} 
\institute{All authors work at \at Universit\'e C\^ote d'Azur (see acknowledgements for details), \email{fmillour@oca.eu}.}
\maketitle

\abstract{
CubeSats are tiny satellites with increasing capabilities. They have been used for more than a decade by universities to train students on space technologies, in a hands-on project aiming at building, launching and operating a real satellite. Still today, one shortcoming of CubeSats is their poor ability to transmit large amounts of data to the ground. A possible way to overcome this limitation relies on optical communications. Universit\'e C\^ote d'Azur is studying the feasibility of a student's CubeSat whose main goal is to transmit data with an optical link to the ground at the moderate rate of 1\,kb/s (or better). In this paper, we will present the current state of the project and its future developments.
}

\section{Introduction, mission objectives}
\label{sec:intro}

CubeSats are small satellites (``SmallSats'' class) made of 10\,cm-side cubes that form a ``unit'' (or a 1U CubeSat). A large number of units can be combined, although the bulk of current developments range from 1U (e.g. Robusta1B), 2U (Spacecube, X-cubesat), up to 3U (e.g. Picsat, Eyesat, NIMPH \cite{fernandez:hal-01274184}). However today, we can start to see 12U CubeSat projects under development by several universities (for example the Grenoble/Toulouse project ATISE \cite{lecoarer:hal-01401693}). These SmallSats are getting more and more attention from the universities because they offer the possibility to teach space-related techniques to students on a hands-on experiment, with a budget that can be reached by a medium-size university.

CubeSats also get more and more attention from companies (e.g. Nexeya\footnote{\url{https://www.nexeyaonline.com/small-sats-satellite-platforms}}, Planet Labs\footnote{\url{https://www.planet.com}}) because they offer fast development cycles for new technologies with reduced costs compared to more ``traditional'' satellites. The drawbacks are an extremely small payload volume and mass, a lack of redundancy, and a perfectible reliability, that can be mitigated by payload miniaturization and the use of satellite constellations, or ``flocks'' \cite{boshuizen2014results}.

The consequence of this tiny size is the limited data transmission capacities that can be integrated into a CubeSat: most of the radio transmitters (67\% of the 1630 radio emitter-receivers included in CubeSats\footnote{according to \url{http://www.nanosats.eu/index.html\#figures}, consulted in June 2018}) use the amateur radio frequencies to transmit data (UHF -- 437\,MHz \& VHF -- 146\,MHz), a few (6\%) use S-band (2.2-3.4\,GHz), some (25\%) use X-band (10\,GHz), and the remaining 2\% use other frequencies. 
The use of radio-frequencies to transmit data from the satellite to the ground has some drawbacks, like the crowding of used frequencies (potentially producing interference), or the poor directivity of the radio beam (enabling hacking of the data reception), not to say the poor data rate of UHF and VHF (about 1\,kb/s). In addition, considering the small available volume in the satellite, the UHF/VHF antennas have to be mechanically deployed, which is adding a risk to the success of the mission. Mitigating this risk by finding antenna schemes robust to deployment failure is an interesting track to look for.

To alleviate this poor data rate, an optical transmission chain (light source -- beam launcher -- telescope -- photodiode) can be considered instead of a radiofrequency chain (transmitter -- TX antenna -- RX antenna -- receiver). An optical transmission chain has some advantages over a radio chain: it has a high directivity, making it difficult to intercept, there is no need to allocate a frequency, and there is a potential to have a high-speed data link (several hundred of Mb/s \cite{Janson2013})

\section{Mission description}
\label{sec:descr}

The main goal of the Nice cube mission (Nice$^3$) is to establish a data optical link between the satellite and the ground, while keeping it in a 1U CubeSat format.

The second goal of the mission is to demonstrate a high-enough transmission rate (higher than 1\,Kb/s).

All the characteristics of the mission are derived from these two key aspects.
Several CubeSat missions have already had a similar goal to produce a data optical link: some (e.g. FitSat \cite{Tanaka2013} or Equisat\footnote{\url{https://brownspace.org}}) use arrays of LEDs to render the CubeSat visible from Earth with a small telescope or even the naked eye. They may communicate with the ground via Morse code \cite{TANAKA2015112}. Other missions (OCSD \cite{Janson2013}, Node \cite{Clements2016}) embed a high-power LASER that is precisely pointed at the ground station. Finally, the C3PO \cite{quintana2016, dhumieres2017} project, and other developments at the US Navy \cite{Rabinovich2004, Peter2010} have the objective of developing the Multiple Quantum Well technology (MQW) to produce a retro-reflecting modulator that can be embedded into a CubeSat. MQW technology allows one to establish an asymmetric optical link with a very high bandwidth.

For Nice$^3$, a first assessment of technologies and available resources led us to favor retro-reflecting solutions that we will present in this paper. Other options will remain possible if retro-reflecting solutions do not converge fast-enough to a mature state (i.e. both space and ground segments are working).

\section{Mission constraints}
\label{sec:constraints}

The first and main constraint of the mission is that it must fit into a 1U CubeSat. This constraint limits the available electrical power onboard and the payload space inside the satellite.


The second constraint, resulting from the main goal of the mission, is to establish a \emph{successful} optical link between the satellite and the ground.

Remembering that the light source is aimed at the satellite from the ground (LASER), this means that the satellite must always present a face with the modulating retro-reflector to the ground station when flying over it, with a precision  to be determined (but anyway better than 10$^\circ$).
This can be achieved either with covering each 6 cube face with a retro-reflector (as in the design presented in Fig.~\ref{fig:cubesat}), or using a passive or an active attitude control system, that we plan to study in details during the project.
At the same time, the satellite must stay in orbit long-enough to fulfil the main mission objectives, and it must be close-enough and have a large-enough reflection surface to establish the optical link with sufficient margins with the ground station.

The third constrain is the compliance to the LOS (Loi des Op\'erations Spatiales\footnote{\url{https://www.legifrance.gouv.fr/affichTexte.do?cidTexte=JORFTEXT000018931380}}), i.e. the satellite must de-orbit back to Earth in less than 25 years.

These three constraints are somewhat contradictory and we will find the optimum values for the orbit altitude in the coming months. The second and last constraints are being tested right now with the STELA\footnote{available at \url{https://logiciels.cnes.fr/fr/content/stela}} tool from CNES (see Table.~\ref{fig:orbit}), and an orbit ranging from 500 to 650\,km seem to be relevant for this mission in order to both comply with the LOS and a typical mission duration of 1 year (in order to allow us some time to set up the satellite in flight, verify its good health, acquire the satellite with the optical ground station, and then perform the data transmission test itself).

\begin{table}[htbp]
\caption{Set of possible orbits for the CubeSat mission that satisfy the LOS. A first range of possible orbits for Nice$^3$ is between 500 and 650\,km with a small eccentricity.}
\label{fig:orbit}      
\includegraphics[width=1\textwidth]{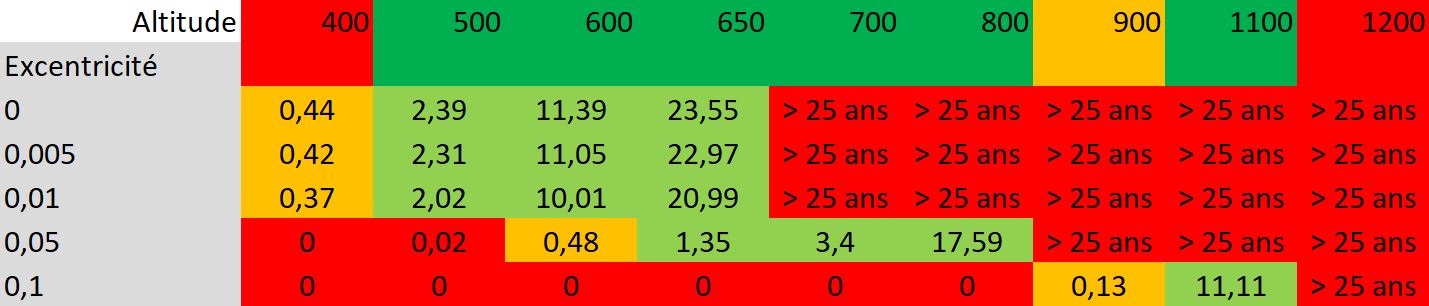}
\end{table}

\section{First sets of definitions of the mission}
\label{sec:def}

The satellite mission contains both a ground station and command center, and the satellite itself. The satellite can be decomposed into a payload and a platform.

\subsection{Description of the ground segment}
\label{sec:GS}

The ground segment will be composed of 3 items: A command and control center, a radio ground station (probably UHF/VHF) for housekeeping telemetry, satellite remote control, and satellite position downlink, and the optical ground station itself, made of a fast-steering telescope and a lasercom setup.

The UHF/VHF ground station will be mounted with on-the-shelf hardware to provide the necessary two-way communications with the satellite. This main radio station will be designed as the project advances.

\begin{figure}[htbp]
	\sidecaption[t]
    \centering
\includegraphics[width=.6\textwidth]{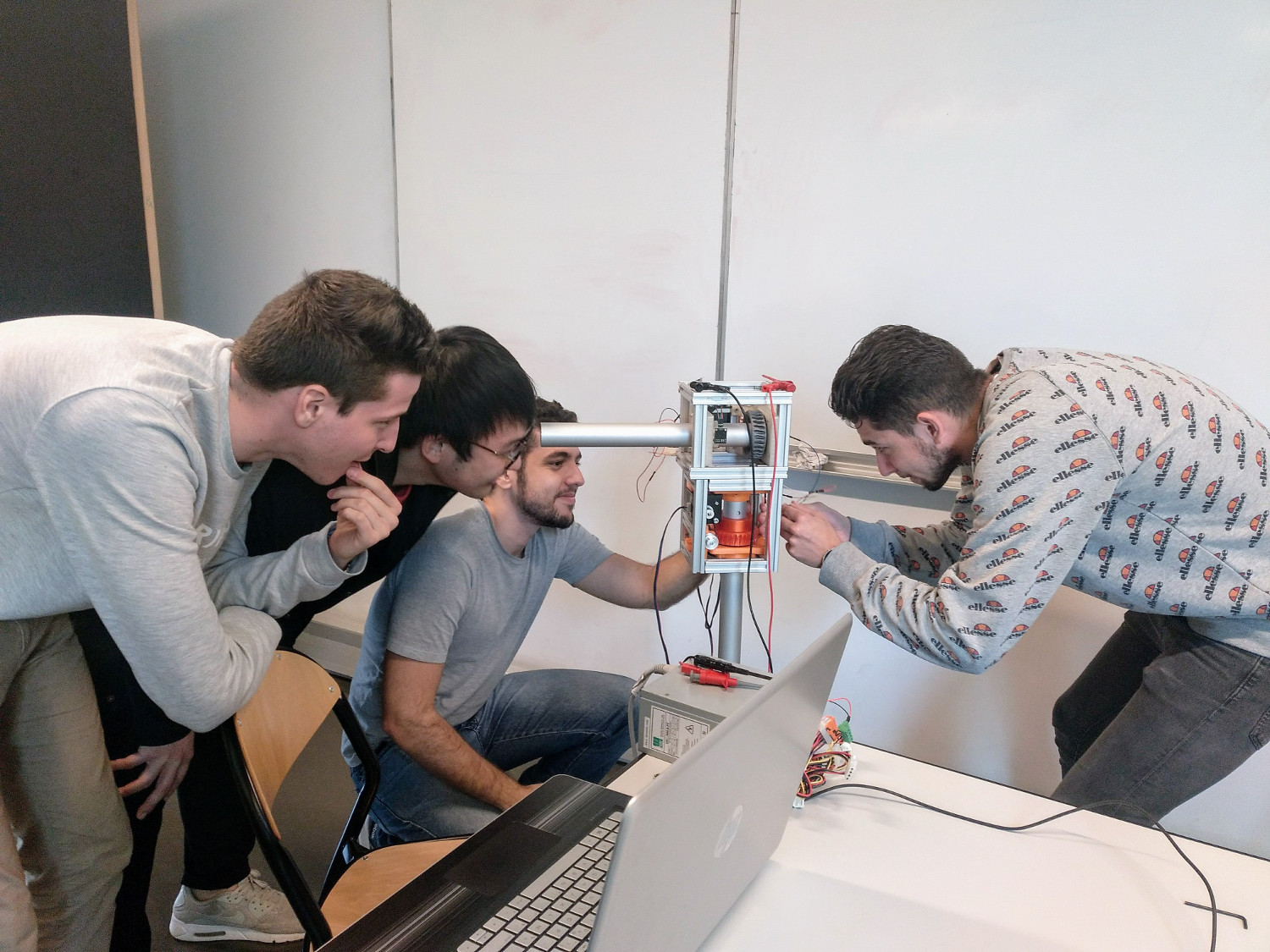}
\includegraphics[height=.45\textwidth]{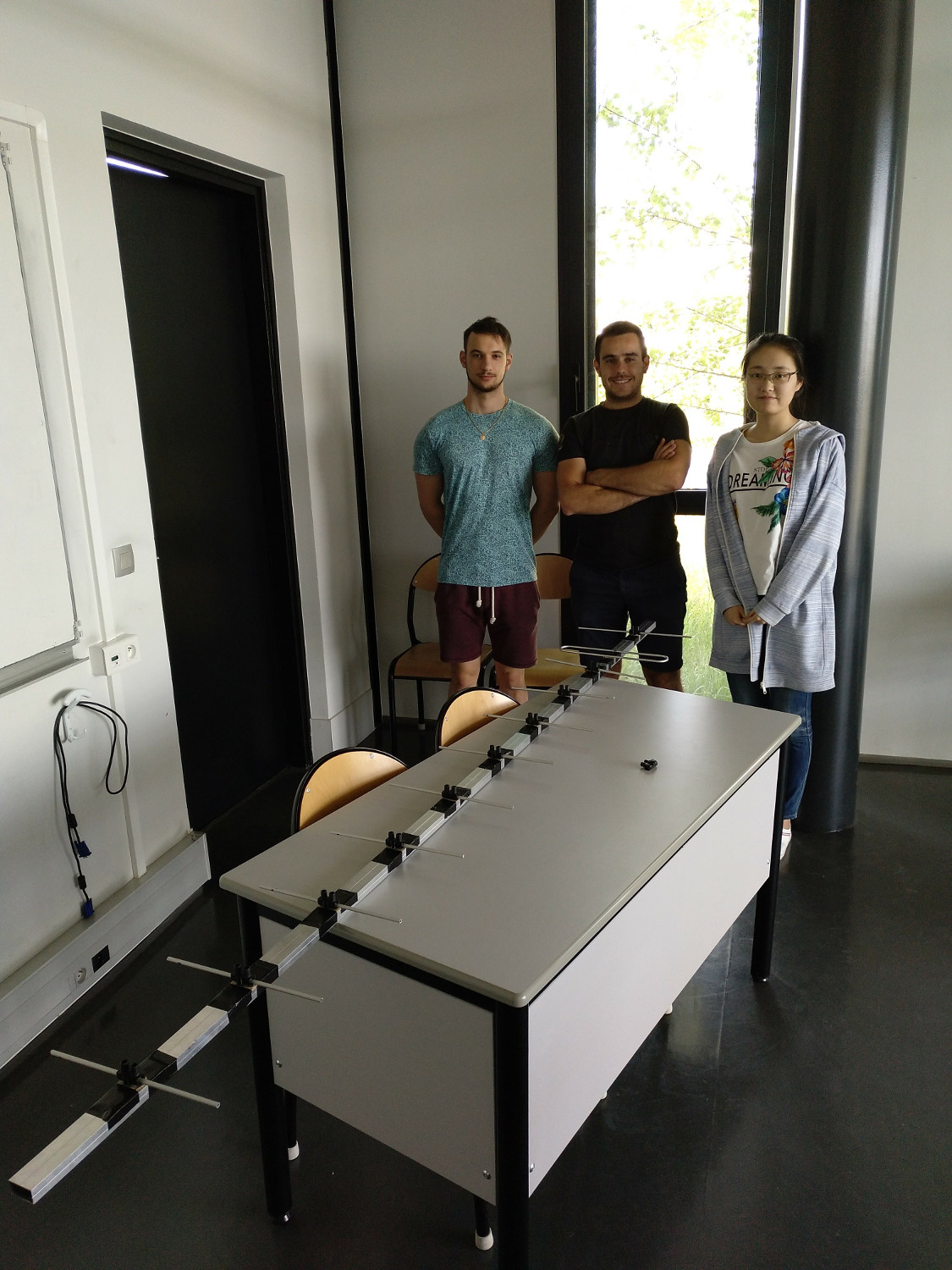}
\caption{Current state of the SATNOGS ground station at Polytech Nice engineering school. On the left side, one can see a group of students working on the control electronics of the mount, and on the right side, one can see the Yagi-Uda antenna being designed by another group of students.}
\label{fig:GS}      
\end{figure}

In the meantime, we started building a UHF SATellite Networked Open Ground Station (SATNOGS)\footnote{\url{https://satnogs.org}} for demonstration purposes with the students of the Polytech Nice-Sophia Antipolis engineering school. The current state of this station is the following: some of the mechanical parts were printed on a 3D printer and assembled, the other parts of the assembly (trusses) being cut from off-the-shelf components. Then, the command electronics was tested and a PCB was designed to integrate an arduino, stepper motors control, endstops control, current sensor and temperature sensors . A first command code was also produced by the students (Fig.~\ref{fig:GS}, left). On the other hand, a Yagi-Uda antenna tuned to 450\,MHz was designed and tested (Fig.~\ref{fig:GS}, right). The next steps are to integrate all the necessary parts (mount, antenna, SDR, arduino, raspberry pi), focus on a 437\,MHz antenna, add a GPS, inertial measurement units, make it battery-operated, to provide a transportable ground station that can be easily demonstrated on conferences or shows.

To give an idea of the dimensions of the optical ground station, a first assessment of the light propagation was made, considering a 1\,W LASER at 1550\,nm wavelength, atmosphere disturbance (seeing) of 3'', light diffraction, atmosphere absorption, etc. This assessment led us to consider a corner cube assembly with a 3\,cm aperture on the satellite, a 20\,cm aperture on the beam launching device (upwards LASER), and a 1.5\,m telescope size for the reception of the reflected signal (downwards-reflected LASER). A typical signal-to-noise ratio of 6 (i.e. $\approx$40 photons per cycle) may be achieved for a 100\,kHz light modulation, giving spacious margins to achieve a data rate of 1\,kb/s.

Note here that the corner cube return beam is aimed back directly, whatever the angle the cube makes relative to the ground station (as long as it is in the 10$^\circ$ misalignment range mentioned above).

All these figures need to be confirmed, but they give a first idea of the typical features of the mission. The MeO 1.5\,m LASER telemetry telescope, located in the Calern plateau less than 50\,km from Nice, and operated by the C\^ote d'Azur Observatory, is a prime candidate to serve as the optical ground station for Nice$^3$.

\subsection{Description of the platform}
\label{sec:platform}

The platform will be comprised of:
\begin{itemize}
\item a mechanical structure holding the necessary electronics and the payload,
\item an electrical procurement system composed of:
\begin{itemize}
\item solar generators (solar panels),
\item energy storage (rechargeable batteries),
\item a power supply unit (PSU),
\end{itemize}
\item an on-board computer (OBC)
\item a radio-communications system composed of:
\begin{itemize}
\item a radio emitter-receiver,
\item an antenna,
\end{itemize}
\item and finally a thermal regulation system.
\end{itemize}

The platform may be built using commercial off the shelf elements from a well-known supplier, but we also investigate the possibility to collaborate with another CSU that developed all these elements in-house, like e.g. the Montpellier University CSU.

\subsection{Description of the payload}
\label{sec:payload}

The Nice$^3$ payload is composed of a retro-reflector, allowing the satellite to send back an optical beam to the emitter on the ground, and an optical modulator, which will encode the data meant to be sent from the satellite to the ground. The optical modulator is the centrepiece of the project, and we are investigating the possible options to achieve the necessary bandwidth of the mission ($\geq 1$\,Kb/s).

\begin{figure}[htbp]
	\sidecaption[t]
    \centering
\includegraphics[width=1.\textwidth]{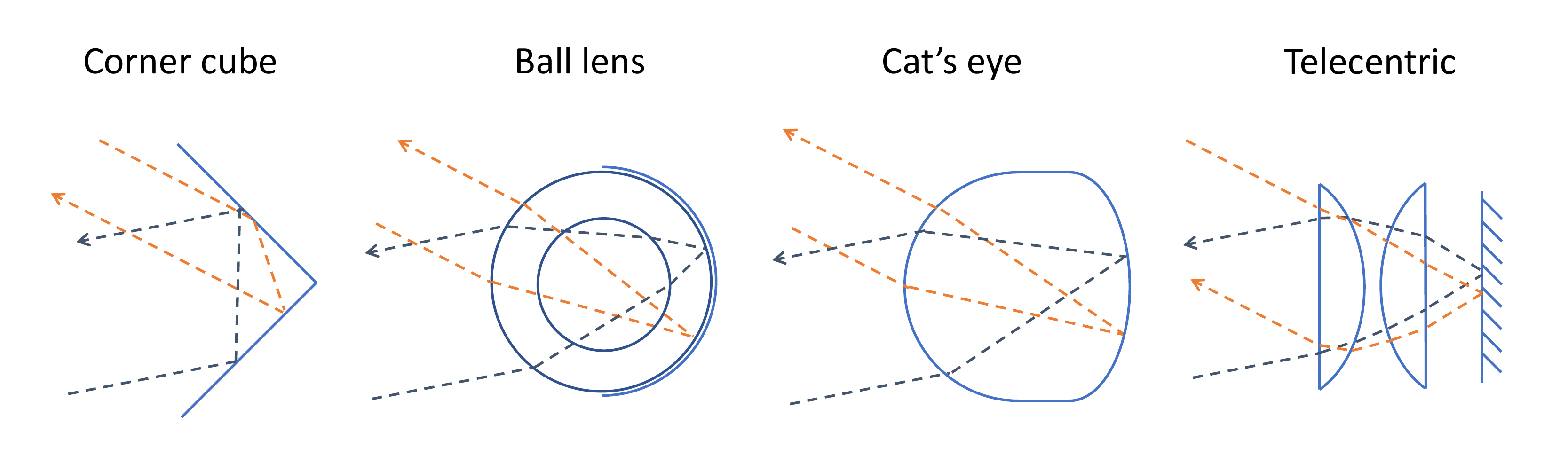}
\caption{Different types of retro-reflectors that will be considered for the mission.}
\label{fig:retro}      
\end{figure}

Retro-reflectors can come in several forms: prismatic retro-reflector (corner cubes) ; hollow corner cubes ; ball lenses ; cat's eyes ; and finally telecentric reflectors (See Figure~\ref{fig:retro}). We are building an optical bench to get measured characteristics of these different types of retro-reflectors, especially the above-mentionned 10$^\circ$ tolerance to misalignment. Once they have been characterized, we will select one type of reflector for the mission, based on their acceptance angle, overall reflectivity, optical quality, etc.

We consider that a retro-reflector with an aperture larger than 3\,cm cannot fit in the satellite, so this is our maximum size.

Optical modulators, which act here as optical shutters, can come in several flavours too: Liquid crystal devices, being in transmission (LCD shutter from Thorlabs) or in reflection (LCOS from Hamamatsu) ; Texas Instrument's Digital Light Processor (DLP) ; Boston Micromachine's modulating reflector (MRR) ; tip/tilt mirrors mounted on piezoelectric actuators (made by Cedrat Technologies and flown in PicSat) ; and finally Multiple Quantum Well technology (MQW) developed by Arianegroup and the US Navy. These different devices have different response times that are listed in Table~\ref{tab:modulcarac}

\begin{table}[htbp]
\caption{\label{tab:modulcarac} Optical modulators considered in the project and their typical characteristics.}
\begin{center}
\begin{tabular}{|l|c|c|c|c|c|}
\hline
\hline
Modulator & Active area & Max. frequency & Power & Mass & space-ready? \\
\hline
LCD & 20 x 20\,mm & 2\,kHz & 300\,mW & 100\,g incl. mount & No \\
LCOS & 16 x 12\,mm & 120\,Hz & 35\,W\footnotemark & - & No \\
DLP & 10 x 6\,mm & 4KHz & 91\,mW & - & No \\
MRR & $\diameter$14\,mm & 200\,kHz & 10\,$\mu$W & 300\,g incl. mount & No \\
Piezo actuator & - & 10\,kHz & 0.75\,W & 12\,g without mirror & Yes \\
MQW  & - & $\geq$10\,MHz & - & - & - \\
\hline
\hline
\end{tabular}
\end{center}
\end{table}

\footnotetext{To be verified.}

We have three possible optical configurations depending on the type of modulator and the type of retro-reflector chosen for the mission.

Depending on choices made on the way the mission is designed (attitude control vs. no attitude control), we will have two different satellite configurations: one without attitude control and 6 modulating retro-reflectors (1 per cube face), and one with an attitude control and just 1 retro-reflector.

The payload may also include a GNSS receiver (e.g. GPS or GALILEO) to locate the satellite in real time by sending its position to the Ground via a radio link, and a high-power LED in order to locate the satellite even when it is not illuminated by the Sun.

\begin{figure}[htbp]
	\sidecaption[t]
    \centering
\includegraphics[width=1.\textwidth]{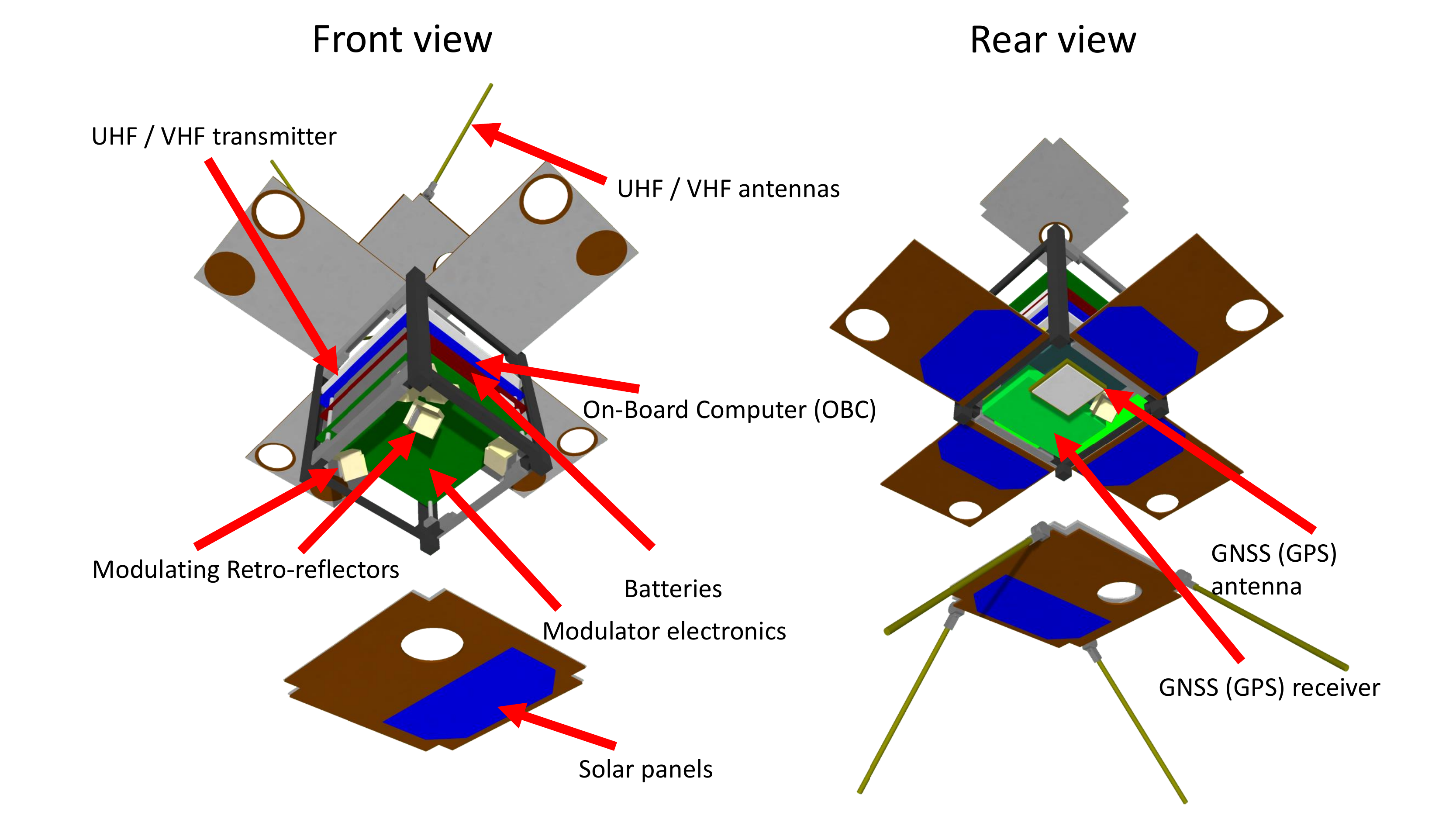}
\caption{Preliminary design of a possible satellite fulfilling several aspects of the mission.}
\label{fig:cubesat}      
\end{figure}

Based on these elements, we designed a first version of the Nice$^3$ satellite, using only off the shelf elements, and hollow corner cubes (see Fig.~\ref{fig:cubesat}).

In this design, the light modulators are embedded in one of the corner cube faces.
No attitude control was considered for this first design, leading us to place one corner cube on each of the satellite face.
We placed an additional generic PC104 electronics board on the satellite stack to figure the modulators control electronics.
We also placed a GNSS receiver and its antenna, in order to send the satellite position to the ground via radio link.

Drilling a hole on each satellite face to let the corner cube go through lead to the necessity to remove one solar cell from the solar panels. The consecutive deficit of power (0.5\,W instead of 1\,W typical) may be mitigated by using solar cells with a different shape, or deployable solar generators.

\begin{figure}[htbp]
    \centering
\includegraphics[width=1.\textwidth]{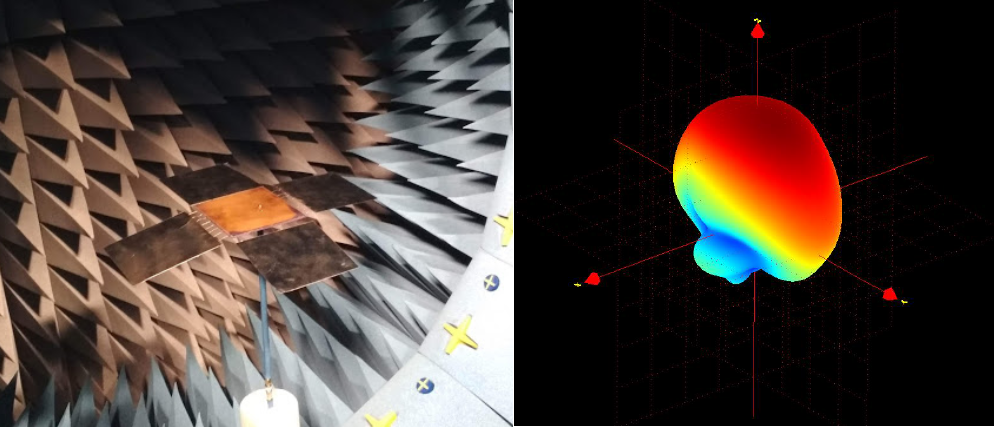}
\caption{{\bf Left:} microstrip patch deployable antenna prototype on the characterization bench. {\bf Right:} measured 3D radiation pattern at 900MHz of this antenna.}
\label{fig:antenna}      
\end{figure}

In this preliminary phase, the integration of a radiolink based on Low-Power Wide Area Network (LP-WAN) has been also investigated. A wireless link over 702km has been already demonstrated at 868MHz using a sounding balloon and LoRa technology \cite{TTN2017}. Several geometries have been proposed for cubesat UHF radio links, including Dipole, Yagi, helical or parabolic structures. All these solution require a mechanical deployment to enable the RF communication, which can be risky. We involved students from Polytech Nice Sophia Antipolis to design a custom antenna based on a microstrip patch using one face of the cubesat, and with four deployable panels. 
This antenna was manufactured by the students (see Fig.~\ref{fig:antenna}) and characterized in a Starlab station. A realized peak gain of 5.4\,dBi was achieved. This solution, using a patch antenna, is mitigating the mechanical deployment risk, as a gain of already 4\,dBi will be obtained in case of deployment failure.
This $\approx$900\,MHz antenna is a first step. The next step will include the design of an antenna in the radio amateur bands.

\section{Conclusion}
\label{sec:conclu}

We are about to finish the phase 0 of the Nice$^3$ cubesat mission. 
During these first 6 months, students worked on the project and brought significant progress to our understanding of the context and difficulties of building a satellite.
We have today a set of first boundaries of the satellite mission. This will enable us to progress further in the mission specifications in the coming months.

Students working on this project come from many horizons. They may come from university masters, like MAUCA, but also from engineering schools, like Polytech Nice Sophia Antipolis, and of course from other formations, like optics BTS. The acknowledgements below list further the formations that follow closely the project.
Making a satellite project with students is an exciting experience and we are preparing for the end of the phase 0 with enthusiasm.

\subsection*{Acknowledgements}

\emph{This project is supported by the C\^ote d'Azur University (UCA), the C\^ote d'Azur Observatory (OCA) and the Centre National des \'Etudes Spatiales (CNES). }

\emph{The Centre Spatial Universitaire (CSU) of the C\^ote d'Azur University (UCA) offers students practical training as part of courses from the Astrophysics master (MAUCA) or the Geophysics master course (master 3G) of UCA, from the Mines Paristech and Polytech Nice-Sophia Antipolis engineering schools, as well as professional experience in research laboratories and institutes Lagrange, Geoazur, LEAT, I3S, CEMEF, Inphyni and INRIA.}

\emph{The authors would like to thank the joint laboratory between Université\'e C\^ote d’Azur, CNRS and Orange, ``Centre de Recherche Mutualis\'e pour les Antennes" (CREMANT), for its support in the microstrip patch antenna characterization.}

%
%

\bibliographystyle{unsrt}
\bibliography{biblio}

\end{document}